\documentclass[aps,twocolumn,showpacs,floatfix]{revtex4}
\usepackage[T1]{fontenc}
\usepackage{graphicx}
\newcommand\forget[1]{}

\begin{document}


\title{Observation of slow light in the noise spectrum of a vertical external cavity surface emitting laser}



\author{A. El Amili$^1$}
\email{karim.el-amili@u-psud.fr}
\author{B.-X. Miranda$^{1,2}$}
\author{F. Goldfarb$^1$}
\author{G. Baili$^3$}
\author{G. Beaudoin$^4$}
\author{I. Sagnes$^4$}
\author{F. Bretenaker$^1$}
\author{M. Alouini$^{2,3}$}
\affiliation{$^1$Laboratoire Aimé Cotton, CNRS-Université Paris Sud 11, 91405 Orsay Cedex, France}
\affiliation{$^2$Institut de Physique de Rennes, CNRS-Université de Rennes I, 35042 Rennes Cedex, France}
\affiliation{$^3$Thales Research and Technology, Campus Polytechnique, 91127 Palaiseau Cedex, France}
\affiliation{$^4$Laboratoire de Photonique et Nanostructures, CNRS, Route de Nozay, 91460 Marcoussis, France}

\date{\today}
\begin{abstract}
The role of coherent population oscillations is evidenced in the noise spectrum of an ultra-low noise lasers. This effect is isolated in the intensity noise spectrum of an optimized single-frequency vertical external cavity surface emitting laser. The coherent population oscillations induced by the lasing mode manifest themselves through their associated dispersion that leads to slow light effects probed by the spontaneous emission present in the non-lasing side modes.
\end{abstract}
\pacs{42.55.Ah, 42.60.Lh, 42.50.Lc}


\maketitle
Since the early works of Sommerfeld \cite{Sommerfeld1} and Brillouin \cite{Brillouin1,Brillouin2} on light propagation through resonant atomic systems, slow and fast light (SFL) have been the subject of considerable research efforts. To control the group velocity of light, various approaches have been proposed and demonstrated, such as, e.~g., electromagnetically induced transparency \cite{Harris1990,Hau}, coherent population oscillations (CPO) \cite{Bigelow1,Bigelow2}, and stimulated Brillouin scattering \cite{Thevenaz}. All these approaches are based on the well known Kramers-Kr\"{o}nig relations stating that a narrow resonance in a given absorption profile gives rise to very strong index dispersion in the medium. Consequently, a pulse of light can propagate through a material slower or faster than the velocity of light in vacuum without violating Einstein's causality \cite{Milonni}. In this framework, the major part of the studies reported in the literature is devoted to single-pass propagation in the considered dispersive medium: the pulse shape or the amplitude modulation of the light is fixed at the entrance of the SFL system. The point is then to investigate how these characteristics evolve during propagation through the medium.

Systems, such as lasers, in which the light is self organized, have not attracted so much attention in this context. Yet, CPO, an ubiquitous mechanism inducing SFL, is present in any active medium provided that a strong optical beam saturates this medium. Thus, CPO must be present in any single frequency laser since the oscillating beam acts as a strong pump which, by definition, saturates the active medium. This effect could be observed using an external probe whose angular frequency is detuned with respect to the oscillating mode, by less than the inverse of the population inversion lifetime $1/\tau_{\mathrm{c}}$. Besides, it has been shown in semiconductor optical amplifiers (SOAs) that CPO induced SFL leads to a significant modification of the spectral noise characteristics at the output of the SOA \cite{Gadi, Perrine}. Consequently, this effect should be also visible in the laser excess noise, using the spontaneous emission present in the non-lasing side longitudinal modes of a single-frequency laser as probe of the CPO effect. To reach this situation, the free spectral range (FSR) of the laser must not be larger than $1/\tau_{\mathrm{c}}$. This is seldom fulfilled in most common lasers. For instance, in ion-doped solid-state lasers, $\tau_{\mathrm{c}}$ is in the range of $1\ \mu$s~-~10~ms \cite{Siegman}. Thus, the FSR of the laser should be smaller than 1~MHz, forbidding single-frequency operation. On the other hand, $\tau_{\mathrm{c}}$ in semiconductor lasers is in the ns range. Consequently, CPO effects are efficient at offset frequencies below a few GHz from the lasing mode \cite{Agrawal}. The FSR of edge emitting semiconductor lasers being around 100 GHz makes them unsuitable for this experiment. However, class-A vertical external cavity surface emitting semiconductor lasers (VECSELs) \cite{Ghaya1} recently developed for their low noise characteristics exhibit i) single-frequency operation, ii) ultra-narrow linewidth \cite{Garnache}, iii) shot-noise limited intensity noise \cite{Ghaya2}, and iv) a FSR in the GHz range. All these characteristics make them perfectly suited for the observation of CPO induced SFL in their noise spectrum.

\begin{figure}[]
      \includegraphics[width=0.8 \linewidth]{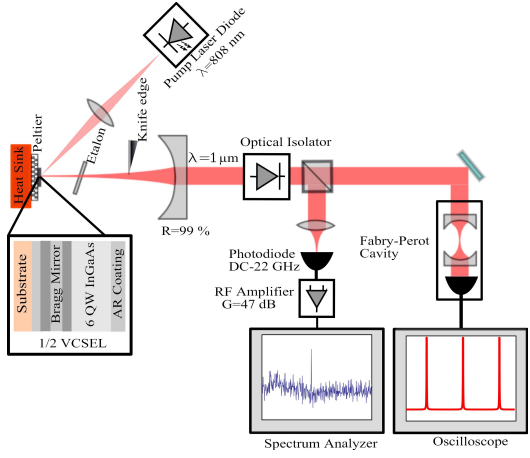}\\
  \caption{Experimental setup. A 808\,nm fibred laser is used to pump the 1/2-VCSEL chip that emits around 1\,$\mu$m. The ex      tended cavity is closed by a 99\% reflection mirror. The étalon is used to obtain the single-frequency regime and a knife edge is inserted to increase the losses in a controllable manner and thus adjust the power inside the cavity. The output light is partly sent to a Fabry-Perot analyzer and partly to a detector followed by an electrical spectrum analyzer.}\label{fig-set-up}
\end{figure}

The laser used in our experiment is a VECSEL which operates at $\sim1\;\mu$m (Fig.\ \ref{fig-set-up}). The 1/2-VCSEL gain chip is a multi-layered stack, over $L_{m}\approx10\,\mu\mathrm{m}$ length, of semiconductors materials. Gain is produced by six InGaAs/GaAsP strained quantum wells grown on a high reflectivity Bragg mirror. The Bragg mirror side is bonded onto a SiC substrate to dissipate the heat towards a Peltier cooler. The top of the gain structure is covered by an anti-reflection coating. The gain is broad ($\sim$ 6 THz bandwidth) and spectrally flat and has been optimized to reach a low threshold \cite{Garnache}. The output mirror (10-cm radius of curvature, 99\ \% reflectivity) is placed at $L \lesssim 10\;\mathrm{cm}$ from the gain structure. In these conditions, $1/2\pi\tau_{\mathrm{c}}$ is not negligible compared with the FSR ($\Delta\gtrsim 1.5\;\mathrm{GHz}$). The laser is optically pumped at 808~nm. The pump is focused to an elliptical spot on the structure with the ellipse aligned with the [110] crystal axis to avoid polarization flips. A $200-\mu$m thick glass étalon is inserted inside the cavity to make the laser single mode. Its spectrum is continuously analyzed with a Fabry-Perot interferometer to ensure that the laser remains monomode and that there is no mode hop during spectra acquisitions.
The noise spectrum is measured using a setup similar to the one described by Baili \textsl{et al.} \cite{Ghaya1}. We use a wide bandwidth photodiode and a low noise radio-frequency amplifier in order to reveal the excess noise due to the beatnotes between the laser line and the spontaneous emission noise at neighboring longitudinal mode frequencies \cite{Ghaya2}. Indeed, the laser output field reads $E\left(t\right)=\sum_{p}\mathcal{A}_{p}e^{-2i\pi\nu_{p} t}+\mathrm{c.c.}$, where $p$ holds for the different mode orders of amplitudes $\mathcal{A}_{p}$ at the cold cavity frequencies $\nu_{p}=\nu_{0}+p~\Delta$. $p=0$ corresponds to the lasing mode, and $p=\pm 1$ to the two closest non-lasing modes, etc... This field leads to the following photocurrent at the output of the detector:
\[i_{ph}\left(t\right)\propto \left|\mathcal{A}_{0}\right|^2+ \sum_{p\neq0}\left|\mathcal{A}_{p}\right|^2+\sum_{p\neq0}\left[\mathcal{A}_{0}\mathcal{A}^{\ast}_{p}
\exp\left(-2i\pi f_{p}t\right)+\mathrm{c.c.}\right],\]
where the side mode fields $\left|\mathcal{A}_{p}\right|$ (containing only spontaneous emission) are very small compared with the lasing mode field $\left|\mathcal{A}_{0}\right|$. Thus, the excess intensity noise, characterized by $\mathcal{A}_{0}\mathcal{A}^{\ast}_{p}$, consists of peaks located at $\left|f_{p}\right|=\left|\nu_{0}-\nu_{p}\right|$ in the Fourier space.
\begin{figure}[]
  \includegraphics[width=0.8\linewidth]{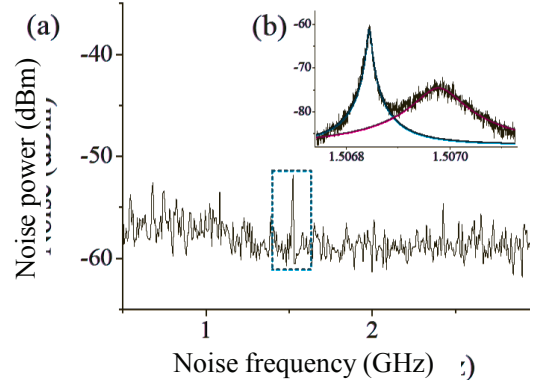}
  \caption{Typical laser intensity noise spectrum. For a cavity length $L\approx$ 10 cm, the beat note frequency appear at the first harmonic of the resonator FSR $\Delta\approx$ 1.5 GHz. The inset is a zoom of the excess noise in the region around $\Delta$. The fact that this noise is composed of two Lorentzian peaks is the signature of a CPO induced gain modulation, that leads to a dispersion effect probed by the non-lasing modes located $\pm\Delta$ from the lasing frequency $\nu_0$.}
  \label{fig-2}
\end{figure}
On that account, the beat frequencies $\left|f_{p}\right|$ occur at harmonics of the FSR in the noise spectrum (Fig.\ \ref{fig-2}). Just above threshold ($\eta-1\ll 1$, where $\eta$ is the laser excitation ratio), the excess noise peak exhibits a Lorentzian shape with a width completely described by the excess of losses $\delta\gamma_{p}$ induced by the étalon on the $p^{\mathrm{th}}$ side mode \cite{Ghaya1}. At the $p^{\mathrm{th}}$ FSR frequency $p\Delta$, the noise spectrum is thus the sum of two Lorentzian peaks due to the beat notes of the lasing mode with the corresponding sidebands ($p^{\mathrm{th}}$ and $-p^{\mathrm{th}}$ modes).
By contrast, when the pumping rate is increased, we found experimentally that the excess noise consists of two peaks separated by $\delta f=f_{p}-f_{-p}\sim$ 100 kHz (inset of Fig.\ \ref{fig-2}). This frequency shift is given by
\begin{equation}
\delta f\approx \nu_{0}\frac{L_{m}}{L+n_{0}L_{m}}(\delta n_p + \delta n_{-p}) , \label{eqdeltaf}
\end{equation}
where $n_{0}$ is the bulk refractive index of the semiconductor structure. $\delta n_{\pm p}$ are the modifications of the refractive index of the structure experienced by the $\pm p$ side modes and induced by the dispersion associated with the CPO effect. In a semiconductor active medium, thanks to the Bogatov effect \cite{Bogatov}, the dispersion is not an odd function of the frequency detuning with respect to $\nu_0$. Thus, $\delta n_p \neq -\delta n_{-p}$ and the two beat note frequencies $f_p$ and $f_{-p}$ corresponding to the $p$ and $-p$ modes occur at slightly different frequencies, as evidenced by the double peak of Fig.\ \ref{fig-2}.
\begin{figure}[h]
   \includegraphics[width=0.8\linewidth]{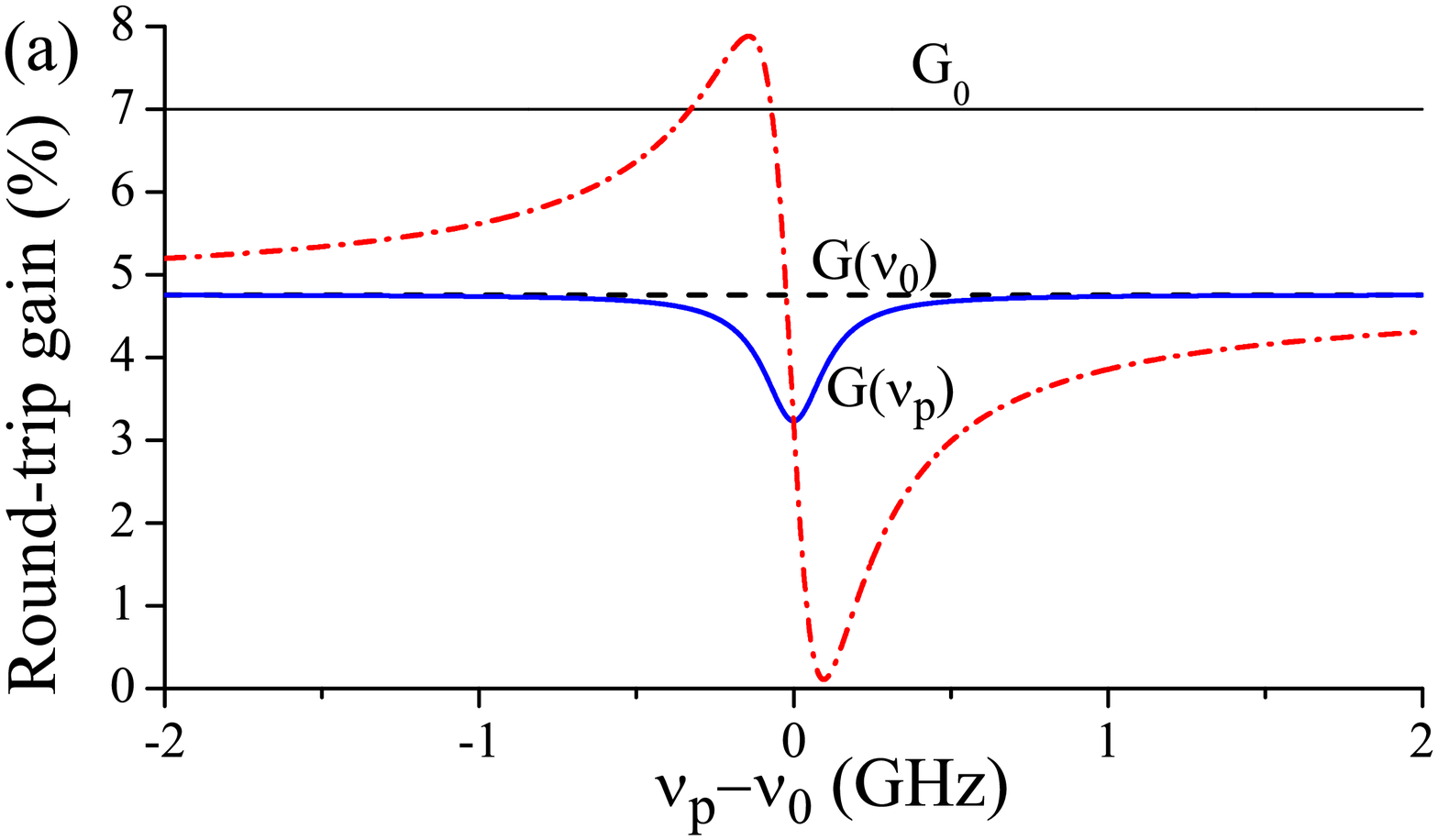}
   \includegraphics[width=0.8\linewidth]{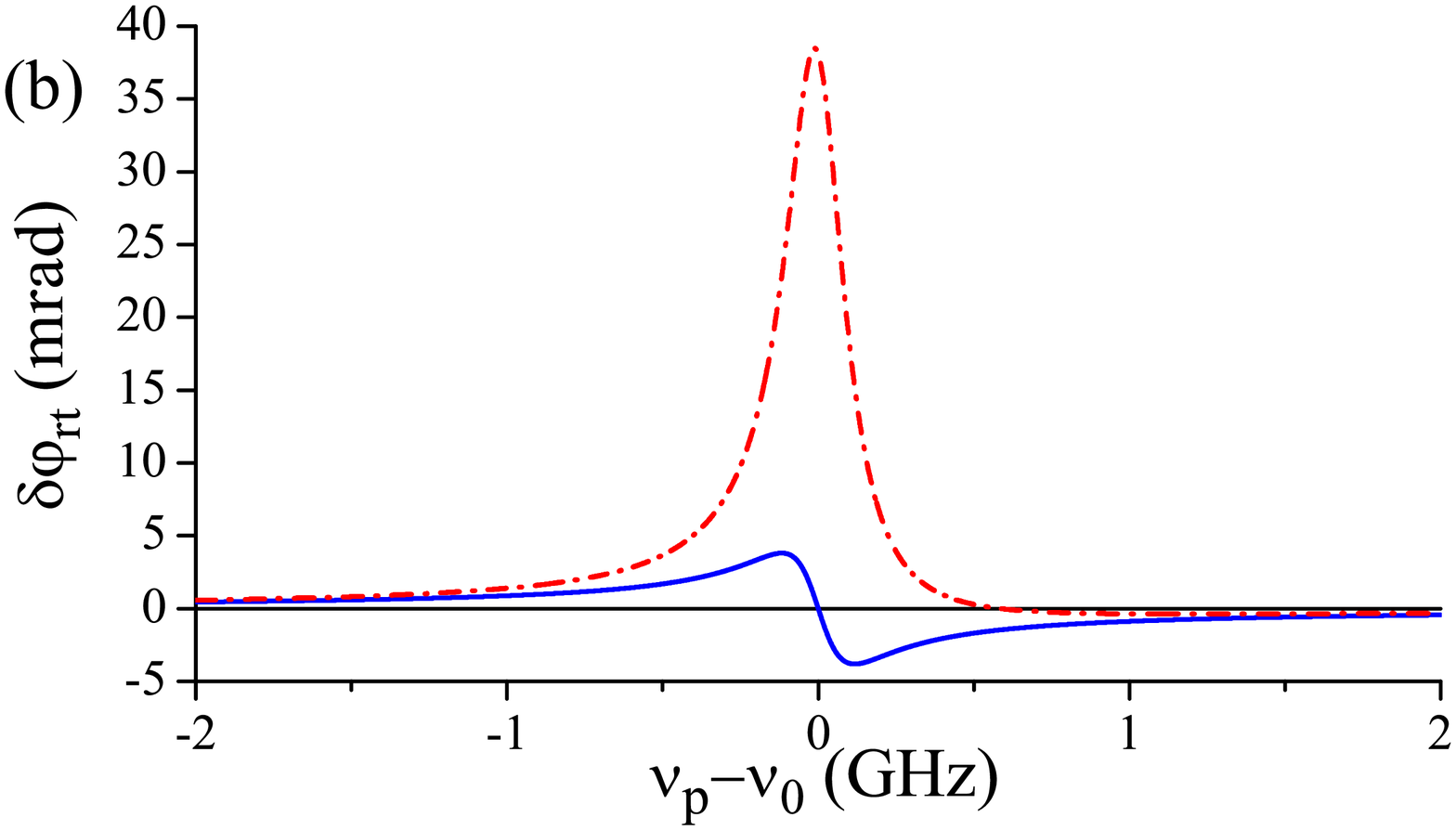}\\
  \caption{(a) Round-trip gain versus probe frequency detuning $\nu_0-\nu_p$. The thin line is the unsaturated gain. The dashed line is the saturated gain for the light at $\nu_0$. The full and dotted-dashed lines are the gains seen by the probe for $\alpha=0$ and $\alpha=5$, respectively. (b) Round-trip phase modification experienced by the side modes for $\alpha=0$ (full line) and $\alpha=5$ (dotted-dashed line). These profiles are plotted from Eqs. (\ref{eqn-gain}) and (\ref{eqn-index}) with $\tau_{\mathrm{c}}=2\ \mathrm{ns}$, $\mathcal{S}=0.5$, and $G_0=2g_0 L_m=0.07$, which correspond to our experimental conditions.}\label{fig-3}
\end{figure}
More precisely, this CPO induced index modification can be derived from the gain medium rate equation. We assume that this medium can be modeled by a two-level system driven by an intracavity light field $E\left(t\right)$ which is the sum of the lasing mode and the two closest side modes:  \[E\left(t\right)=\mathcal{A}_{0}e^{-2i\pi\nu_{0}t}+\mathcal{A}_{-1}e^{-2i\pi\nu_{-1}t}+\mathcal{A}_{1}e^{-2i\pi\nu_{1} t}+\mathrm{c.c.}.\]
As the étalon forces the laser to operate in single mode regime, one has $\left|A_{0}\right|^2\gg\left|A_{-1}\right|^2\approx\left|A_{1}\right|^2$. Consequently, we consider only the beat notes between the lasing and the adjacent modes which create modulations of the population inversion at frequencies close to $\Delta$. Under these assumptions, the gain $g\left(\nu_{p}\right)$ and the refractive index variation $\delta n\left(\nu_{p}\right) = n\left(\nu_{p}\right)-n_0$ seen by the side modes, that can be considered as weak probes, are given by \cite{Agrawal2}:
\begin{eqnarray}
g(\nu_{p}) &=& \frac{g_{0}}{1+\mathcal{S}}\left\{1-\frac{\mathcal{S}\left[(1+\mathcal{S})+\alpha 2\pi\left(\nu_{0}-\nu_{p}\right)\tau_{c}\right]}{(1+\mathcal{S})^2+\left[2\pi\left(\nu_{0}-\nu_{p}\right)  \tau_{c}\right]^2}\right\}, \qquad  \label{eqn-gain} \\
 \delta n\left(\nu_{p}\right)
 &=& \frac{c}{4\pi\nu_{0}}\frac{g_{0}\mathcal{S}}{1+\mathcal{S}}
 \frac{2\pi\left(\nu_{0}-\nu_{p}\right)\tau_{c}+\alpha(1+\mathcal{S})}{\left(1+\mathcal{S}\right)^2+\left[2\pi\left(\nu_{0}-\nu_{p}\right)\tau_{c}\right]^2}\ . \label{eqn-index}
\end{eqnarray}
Here $g_{0}$ is the unsaturated gain and $\mathcal{S}$ the saturation parameter. $\alpha$ is the phase-intensity coupling coefficient (Henry's factor) that is responsible for the Bogatov effect. Eq.\ (\ref{eqn-gain}) describes two phenomena: i) the self-saturation of the gain at $\nu_0$ by the field at $\nu_0$ [dashed line in Fig.\ \ref{fig-3}(a)] and ii) the modifications due to the CPO effect of the gains probed by the side modes at $\nu_{\pm p}$. The evolution of this gain versus probe frequency is plotted as a full line (resp. dotted-dashed line) in Fig.\ \ref{fig-3}(a) for $\alpha=0$ (resp. $\alpha=5$). This CPO effect is also responsible for the modification of the refractive index seen by the side modes which modifies the round-trip phase accumulated by each side mode [see Fig.\ \ref{fig-3}(b)]. With $\alpha\neq 0$, we notice that the phase shifts for two symmetric side modes are not opposite, restraining $\delta f$ from vanishing [see eq.\ (\ref{eqdeltaf})].
\begin{figure}[]
  \includegraphics[width=1.0\linewidth]{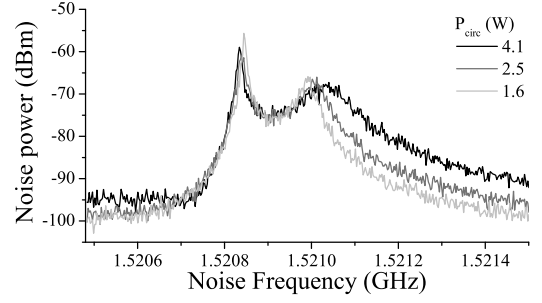}
  \caption{Experimental noise spectrum for different intracavity powers. The difference between the widths of the two peaks is clearly visible. The peak widths and the spacing increase with the intracavity power. Resolution Bandwidth=1 kHz.}
  \label{fig:DoublePeaks}
\end{figure}
Fig.\ \ref{fig:DoublePeaks} shows the double peak for different intracavity powers $P_{\mathrm{circ}}$ (defined as the power of one of the two traveling waves creating the intracavity standing wave). It should be noticed that the two excess noise peak profiles have different widths. This is explained by the fact that at the first order, the widths depend on the losses induced by the intracavity étalon. These extra losses lead to the following extra loss rates for the $p^{\mathrm{th}}$ side mode:
\begin{equation}\label{eqn-T}
\delta\gamma_{p}=2 \Delta \left[1-\mathcal{T}\left(\nu_{p}\right)\right] ,
\end{equation}
where $\mathcal{T}\left(\nu_{p}\right)$ is the étalon intensity transmission for that mode.
\begin{figure}[]
   \includegraphics[width=0.8\linewidth]{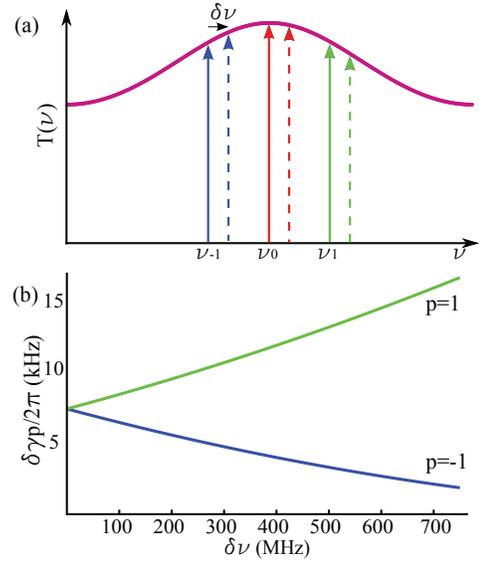}\\
  \caption{(a) Etalon transmission versus frequency. When the lasing mode frequency is shifted by $\delta\nu$ from the maximum of étalon transmission, the transmissions for the side modes at $\nu_{\pm 1}$ are no longer equal. (b) Extra loss rates $\delta\gamma_{\pm 1}$ versus $\delta\nu$.}\label{fig-5}
\end{figure}
When the lasing mode frequency $\nu_{0}$ coincides with a maximum of the transmission spectrum, both side modes transmissions are equal: $\mathcal{T}\left(\nu_{1}\right)=\mathcal{T}\left(\nu_{-1}\right)<1$ and the peak widths are also equal: $\delta\gamma_{1}=\delta\gamma_{-1}$. But if $\nu_{0}$ is shifted by $\delta\nu >0$ from the étalon resonance frequency [see Fig.\ \ref{fig-5}(a)], the étalon transmission for mode $p=+1$ (resp. $p=-1$) decreases (resp. increases) with $\delta\nu$. Figure \ref{fig-5}(b) shows the effect of such a detuning on the extra loss rates $\delta\gamma_{\pm 1}$.
\begin{figure}[]
 \includegraphics[width=0.8\linewidth]{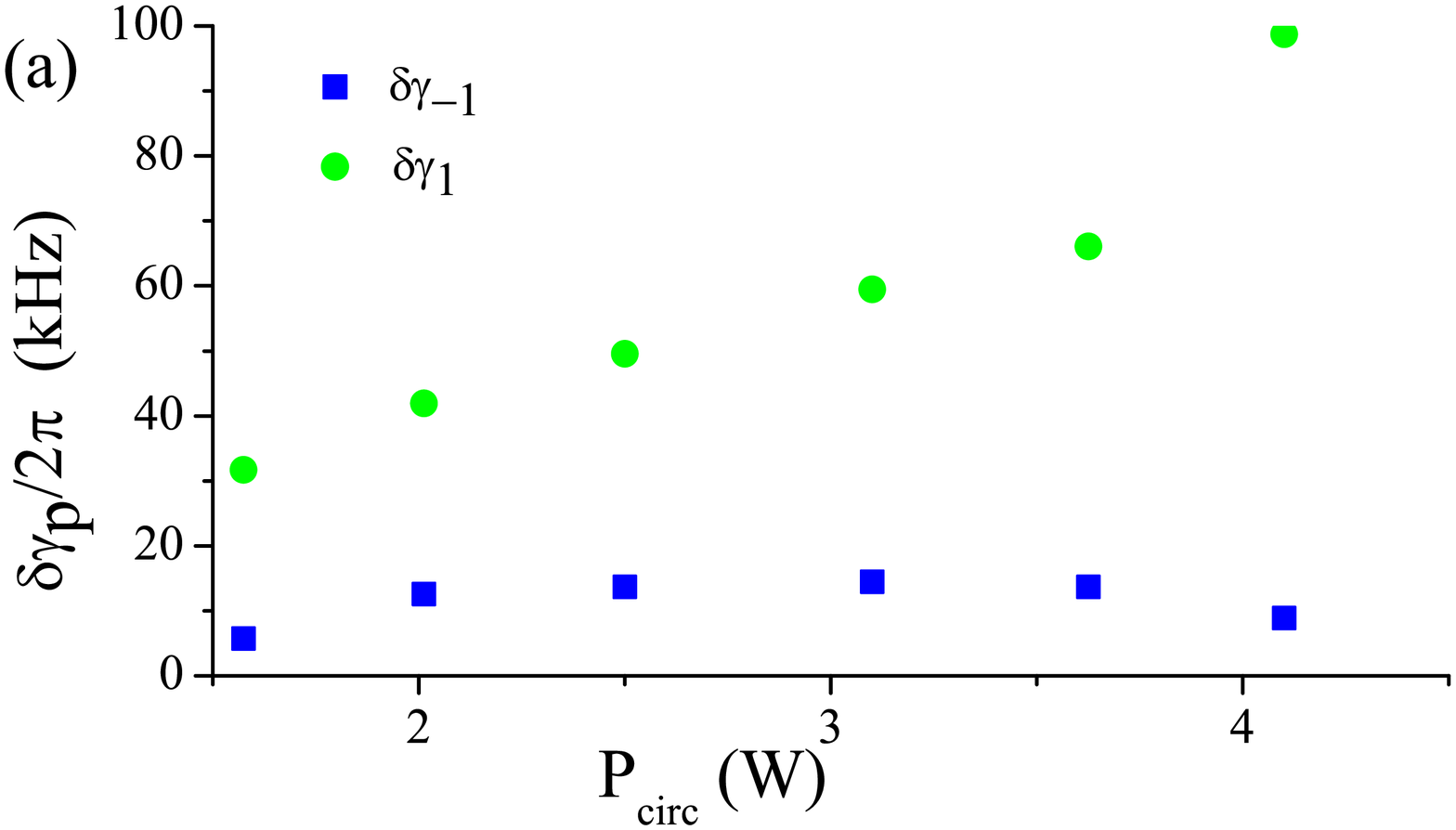}
 \includegraphics[width=0.8\linewidth]{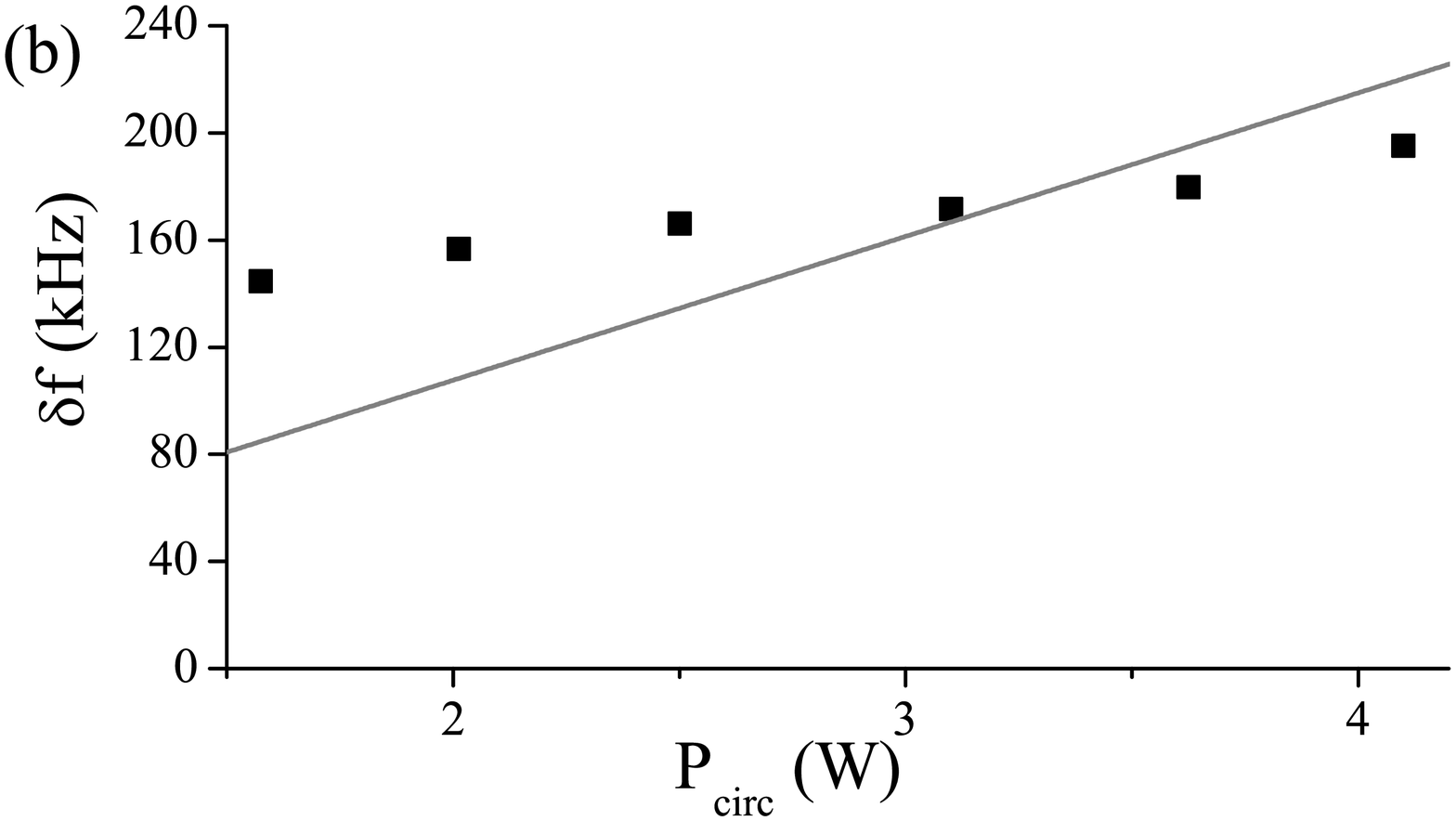}
 \caption{(a) Peak widths $\delta\gamma_{\pm 1}$ versus intracavity power. (b) Peak spacing $\delta f$ versus intracavity power. Squares: measurements. Full line: prediction obtained from eqs. (\ref{eqdeltaf}) and (\ref{eqn-index}) with the same parameters as in Fig.\ \ref{fig-3}}
 \label{fig:WidthAndSpacing}
\end{figure}
We check from the experiment whether this evolution of the extra losses experienced by the side modes correctly explains the widths of the two peaks, like in the simple model of Ref.~\cite{Ghaya2}. The two peaks of Fig.\ \ref{fig:DoublePeaks} are fitted by two Lorentzians in which some asymmetry is included to take into account the Bogatov effect induced by the Henry factor \cite{Bogatov}. Fig.\ \ref{fig:WidthAndSpacing}(a) reproduces the evolution of the peak widths versus intracavity power. This intracavity power is varied by introducing controlled diffraction losses inside the cavity using a knife edge for a constant pump power, in order to keep $g_0$ constant. The variation of the intracavity power modifies the laser frequency shift $\delta\nu$, leading to different evolutions of $\delta\gamma_{1}$ and $\delta\gamma_{-1}$, as expected from Fig.\ \ref{fig-5}(b). However, the magnitudes of the experimentally observed variations of these widths are significantly larger than those calculated in the simple linear model of Eq.\ (\ref{eqn-T}), suggesting the enhancement of this effect by nonlinear contributions. Moreover, it is expected that increasing the intracavity power, and thus the gain saturation, leads to an increase of $\delta f$. Fig.\ \ref{fig:WidthAndSpacing}(b) clearly shows that the frequency shift $\delta f$ between the two peaks increases with the intracavity power, evidencing the nonlinear origin of the double peak noise spectrum expected from Eq.\ (\ref{eqn-index}). The full line in Fig.\ \ref{fig:WidthAndSpacing}(b) is obtained from eqs. (\ref{eqdeltaf}) and (\ref{eqn-index}) with our experimental parameters. It shows that our simple model based on a two-level system including Henry's factor gives the good order of magnitude for $\delta f$ and the correct sign for its evolution versus intracavity power. One should not be surprised by the fact that the agreement with the measurements is not perfect: the model of eqs. (\ref{eqn-gain}) and (\ref{eqn-index}) is too crude to fully describe the gain and index saturation in strained quantum wells. Moreover, we overlooked many effects that may lead to a discrepancy with respect to our simple approach such as i) the variation of $\alpha$ with the carrier density, ii) the thermally induced variations of the index and of the laser mode diameter, iii) the variations of $\tau$ with the carrier density, iv) the possible existence of an offset in $\delta f$ due to the linear dispersion of the gain medium and the étalon. Notice also that since the cavity FSR $\Delta$ is larger than the width of the CPO dip of Fig.\ \ref{fig-3}(a), we are probe the wings of the dispersion profile of Fig.\ \ref{fig-3}(b), i.~e., in the slow light regime. Moreover, we have checked that this phenomenon is not related to a coupled cavity effect since we observed exactly the same behavior of the noise spectrum with another 1/2-VCSEL without any anti-reflection coating. If the splitting between the two peaks were due to a coupled cavity effect, it should be completely different in the absence of the anti-reflection coating, contrary to our observations.

In conclusion, we experimentally evidenced the existence of intracavity slow light effects in a laser induced by the CPO mechanism. These effects are probed by the laser spontaneous emission noise present in the non lasing modes. We have shown that this noise is a very efficient probe to explore the intracavity CPO effects and their evolution with the laser parameters such as the intracavity power. Moreover, we have predicted that this first observation of slow light inside a laser cavity should be able to lead to intracavity fast light if the side mode frequencies are closer to the lasing mode frequencies, i.~e., for a longer cavity. This opens interesting perspectives on the study of intracavity fast light \cite{Gadi,Perrine} which raises numerous interests for applications to sensors \cite{Shariar1,Shariar2}. Moreover, the study of the phase noise of the light present in the side modes of such a laser should lead to interesting features including the noise correlations induced by the laser nonlinear effects.

The authors acknowledge partial support from the Agence Nationale de la Recherche, the Triangle de la Physique, and the Région Bretagne.


\begin{thebibliography}{10}
\expandafter\ifx\csname natexlab\endcsname\relax\def\natexlab#1{#1}\fi
\expandafter\ifx\csname bibnamefont\endcsname\relax
  \def\bibnamefont#1{#1}\fi
\expandafter\ifx\csname bibfnamefont\endcsname\relax
  \def\bibfnamefont#1{#1}\fi
\expandafter\ifx\csname citenamefont\endcsname\relax
  \def\citenamefont#1{#1}\fi
\expandafter\ifx\csname url\endcsname\relax
  \def\url#1{\texttt{#1}}\fi
\expandafter\ifx\csname urlprefix\endcsname\relax\def\urlprefix{URL }\fi
\providecommand{\bibinfo}[2]{#2}
\providecommand{\eprint}[2][]{\url{#2}}


\bibitem{Sommerfeld1} A. Sommerfeld, Ann. Physik \textbf{44}, 177 (1914).

\bibitem{Brillouin1} L. Brillouin, Ann. Physik \textbf{44}, 203 (1914).

\bibitem{Brillouin2} L. Brillouin, \textit{Wave propagation and group velocity} (Academic Press, New York, 1960).

\bibitem{Harris1990} S. E. Harris, J. E. Field, and A. Imamo\u{g}lu, Phys. Rev. Lett. \textbf{64}, 1107 (1990).

\bibitem{Hau} L. V. Hau, S. E. Harris, Z. Dutton, and T. Behroozi, Nature \textbf{397}, 594 (1999).

\bibitem{Bigelow1} M. S. Bigelow, N. N. Lepeshkin, and R. W. Boyd, Phys. Rev. Lett. \textbf{90}, 113903 (2003).

\bibitem{Bigelow2} M. S.  Bigelow, N. N. Lepeshkin, and R. W. Boyd, Science \textbf{200}, 200 (2003).

\bibitem{Thevenaz} L. Thevenaz, Nature Photonics \textbf{2}, 474 (2008).

\bibitem{Milonni} P. W. Milonni, \textit{Fast Light, Slow light, and Left-Handed Light} (Taylor and Francis, New York, 2005).

\bibitem{Gadi} E. Shumakher, S. Ó Dúill, and G. Eisenstein, Opt. Lett. \textbf{34}, 1940 (2009).

\bibitem{Perrine} P. Berger \textit{et al.}, C. R. Physique \textbf{10}, 991 (2009).

\bibitem{Siegman} A. E. Siegman, \textit{Lasers} (University Science Books, Mill Valley, 1986).

\bibitem{Agrawal} G. P. Agrawal and N. K. Dutta, \textit{Semiconductor Lasers, 2nd edition} (Springer, Berlin, 1993).

\bibitem{Ghaya1} G. Baili \textit{et al.}, Opt. Lett. \textbf{31}, 62-64 (2006).

\bibitem{Garnache} A. Laurain \textit{et al.}, Opt. Expr. \textbf{17}, 9503-9508 (2009).

\bibitem{Ghaya2} G. Baili \textit{et al.}, J. Lightwave Technol. \textbf{26}, 8 (2008).

\bibitem{Bogatov} A. P. Bogatov, P. G. Eliseev, and B. N. Sverdlov, IEEE J. Quantum Electron. \textbf{11}, 510-515 (1975).

\bibitem{Agrawal2} G. P. Agrawal, J. Opt. Soc. Am. B \textbf{5}, 147 (1988).

\bibitem{Shariar1} G. S. Pati, M. Salit, K. Salit, and M. S. Shahriar, Phys. Rev. Lett. \textbf{99}, 133601 (2007).

\bibitem{Shariar2} M. S. Shahriar \textit{et al.}, Phys. Rev. A \textbf{75}, 053807 (2007).

\end{thebibliography}
\end{document}